\pdfoutput=1 
\documentclass[12pt]{article}
\usepackage{amsmath,amssymb}
\usepackage{graphicx}
\usepackage{subfigure}
\usepackage[pdftex]{hyperref}
\DeclareGraphicsExtensions{.eps,.bmp,.wmf,.jpg,.pdf}
\numberwithin{equation}{section}
\def\be{\begin{equation}}
\def\ee{\end{equation}}

\def\bea{\begin{eqnarray}}
\def\eea{\end{eqnarray}}

\title{Current observational constraints on holographic dark energy model}
\author{L.N. Granda\thanks{ngranda@univalle.edu.co} ,\, W. Cardona\thanks{wilalbca@univalle.edu.co} \, and\  A. Oliveros\thanks{alexogar@univalle.edu.co}\\
Department of Physics, Universidad del Valle\\ A.A. 25360, Cali,
Colombia} 
\date{}
\begin{document}
\maketitle

\begin{abstract}
\noindent We consider the cosmological constraints on the holographic dark energy model 
by using the data set available from the type Ia supernovae (SNIa), CMB and BAO observations. The constrained parameters are critical to  determine the quintessence or quintom character the model. The SNIa and joint SNIa+CMB+BAO analysis give the best-fit results for $\beta$ with priors on $\Omega_{m0}$ and $\omega_0$. Using montecarlo we obtained the best-fit values for $\beta$, $\Omega_{m0}$ and $\omega_0$. The statefinder and $Om$ diagnosis have been used to characterize the model over other DE models.
\end{abstract}

\section{Introduction}
\noindent 
The astrophysical data from distant Ia supernovae observations \cite{riess}, \cite{kowalski}, cosmic microwave background anisotropy \cite{spergel}, and large scale galaxy surveys \cite{tegmark}, \cite{tegmark2}, all indicate that the current Universe is not only expanding, it is accelerating due to some kind of  negative-pressure form of matter known as dark energy (\cite{copeland},\cite{turner},\cite{sahni},\cite{padmanabhan}). The combined analysis of cosmological observations also suggests that the universe is spatially flat, and  consists of about $\sim 1/3$
of dark matter (the known baryonic and nonbaryonic dark matter), distributed in clustered structures (galaxies, clusters of galaxies, etc.) and $\sim 2/3$ of homogeneously distributed (unclustered) dark energy with negative pressure. Despite the high percentage of the dark energy component, its nature as well as its cosmological origin remain unknown at present and a wide variety of models have been proposed to explain the nature of the dark energy and the accelerated expansion (see \cite{copeland,turner,sahni,padmanabhan} for review).
Among the different models of dark energy, the holographic dark energy approach is quite interesting as it incorporates some concepts of the quantum gravity known as the holographic principle (\cite{bekenstein, thooft, bousso, cohen, susskind}),which first appeared in the context of black holes \cite{thooft} and later extended by Susskind \cite{susskind} to string theory. According to the holographic principle, the entropy of a system scales not with its volume, but with its surface area. In the cosmological context, the holographic principle will set an upper bound on the entropy of the universe \cite{fischler}. In the work \cite{cohen}, it was suggested that in quantum field theory a short distance cut-off is related to a long distance cut-off (infra-red cut-off $L$) due to the limit set by black hole formation, namely, if is the quantum zero-point energy density caused by a short distance cut-off, the total energy in a region of size $L$ should not exceed the mass of a black hole of the same size, thus $L^3\rho_\Lambda\leq LM_p^2$. Applied to the dark energy issue, if we take the whole universe into account, then the vacuum energy related to this holographic principle is viewed as dark energy, usually called holographic dark energy \cite{cohen} \cite{hsu}, \cite{li}. The largest $L$ allowed is the one saturating this inequality so that we get the holographic dark energy density $\rho_\Lambda=3c^2M_p^2L^{-2}$ where $c^2$ is a numerical constant and $M_p^{-2}=8\pi G$.\\

Choosing the Hubble horizon $H^{-1}$ as the infrared cut-off, the resulting $\rho_\Lambda$ is comparable to the observational density of dark energy \cite{horava}, \cite{hsu}. However, in \cite{hsu} it was pointed out that in this case the resulting equation-of state parameter (EoS) is equal to zero, behaving as pressureless matter which cannot give accelerated expansion. The particle horizon \cite{li} also results with an EoS parameter larger than $-1/3$, which is not enough to satisfy the current observational data, but the infrared cut-off given by the future event horizon \cite{li}, yields the desired result of accelerated expansion with an EoS parameter less than $-1/3$, despite the fact that it has problems with the causality. Another holographic DE model have been considered in \cite{sergei}, \cite{sergei1}.\\ 
Based on dimensional arguments and the fact that the time derivative of the Hubble parameter naturally appears in the cosmological equations, in \cite{granda} we have proposed an infrared cut-off for the holographic density of the form $\rho\approx\alpha H^2+\beta\dot{H}$. Though the theoretical root of the holographic dark energy is still unknown, this proposal, which depends on the scale factor and its derivatives, may point in the correct direction as it can describe the dynamics of the late time cosmological evolution in 
a good agreement with the astrophysical observations. Another interesting fact of this model is that the resulting Hubble parameter (and hence the total density) contains a matter and radiation component \cite{granda}, which become relevant  at high redshifts (radiation) in good agreement with the BBN theory, and (the matter component) explains the cosmic coincidence. An important fact is that this model can exhibit quintom nature without the need to introduce any exotic matter. This model has also proven to be useful in the reconstruction of different scalar field models of dark energy which reproduce the late time cosmological dynamics in a way consistent with the observations \cite{granda2,granda3,granda4}.\\
In this paper we study the cosmological constraints on the holographic dark energy model given in \cite{granda} using 
the new dataset of type Ia supernovae (SNIa), called Union compilation. The Union compilation contains 414 SNIa and reduces to 307 SNIa after selection cuts. This 307 SNIa data set complemented with the CMB anisotropy and BAO (baryon acoustic oscillation) data, will be used to fit the parameter $\beta$ of the model. Once $\beta$ is known, it in turn fixes the parameter $\alpha$ and the integration constant via the flatness condition and the current (x=0) equation of state for the dark energy (see Eqs. (\ref{eq9}) and (\ref{eq10}) bellow). In the section 2 we review the main aspects of the model, in section 3 we fit the $\beta$ parameter with the 307 SNIa data set, in section 4 we use the joint SNIa, CMB and BAO data analysis to constraint $\beta$ and show the best-fit $\beta$ for each studied data set. The montecarlo 
simulation is used in section 5 to constraint the three parameters $\Omega_{m0}, \omega_0, \beta$ using the joint SNIa+CMB+BAO analysis, and in section 6 we give the statefinder and $Om$ diagnosis to contrast the studied holographic DE model with the $\Lambda$CDM and other dynamical DE models. 
\section{The Model}
Let us start with the main features of the holographic dark energy model. The holographic dark energy density is given by
\begin{equation}\label{eq2}
\rho_\Lambda=3\left(\alpha H^2+\beta\dot{H}\right)
\end{equation}
\noindent where $\alpha$ and $\beta$ are constants to be determined and $H=\dot{a}/a$ is the Hubble parameter. The usual Friedmann equation is
\begin{equation}\label{eq3}
H^2=\frac{1}{3}\left(\rho_m+\rho_r+\rho_\Lambda\right)
\end{equation}
where we have taken $8\pi G=1$ and $\rho_{m}$, $\rho_{r}$ terms are the contributions of non-relativistic matter and radiation, respectively.
\noindent Setting $x=\ln{a}$, we can rewrite the Friedmann equation as follows
\begin{equation}\label{eq4}
H^2=\frac{1}{3}\left(\rho_{m0}e^{-3x}+\rho_{r0}e^{-4x}\right)+\alpha H^2+\frac{\beta}{2}\frac{dH^2}{dx}
\end{equation}
\noindent Introducing the scaled Hubble expansion rate $\tilde{H}=H/H_0$, where $H_0$ is the present value of the Hubble parameter (for $x=0$), the above Friedman equation becomes
\begin{equation}\label{eq5}
\tilde{H}^2=\Omega_{m0}e^{-3x}+\Omega_{r0}e^{-4x}+\alpha\tilde{H}^2
+\frac{\beta}{2}\frac{d\tilde{H}^2}{dx}
\end{equation}
where $\Omega_{m0}=\rho_{m0}/3H_0^2$ and $\Omega_{r0}=\rho_{r0}/3H_0^2$ are the current density parameters
of non-relativistic matter and radiation. Solving the equation (\ref{eq5}), we obtain
\begin{equation}\label{eq6}
\tilde{H}^2=\frac{2}{3\beta-2\alpha+2}\Omega_{m0}e^{-3x}
+\frac{1}{2\beta-\alpha+1}\Omega_{r0}e^{-4x}+Ce^{-2 x(\alpha-1)/\beta}
\end{equation}
where $C$ is an integration constant. 
From this equation, the following equation for the holographic density is obtained
\begin{equation}\label{eq7}
\tilde{\rho}_{\Lambda}=\frac{3\beta-2\alpha}{2\alpha-3\beta-2}\Omega_{m0}e^{-3x}
+\frac{2\beta-\alpha}{\alpha-2\beta-1}\Omega_{r0}e^{-4x}+Ce^{-2 x(\alpha-1)/\beta}
\end{equation}
with $\tilde{\rho}_{\Lambda}=\frac{\rho_{\Lambda}}{3H_0^2}$. The energy conservation equation gives the corresponding holographic pressure
\begin{equation}\label{eq8}
\tilde{p}_{\Lambda}=\frac{2\alpha-3\beta-2}{3\beta}\,Ce^{-2 x(\alpha-1)/\beta}+\frac{2\beta-\alpha}{3(\alpha-2\beta-1)}\,
\Omega_{r0}e^{-4x}
\end{equation}
There are three constants $\alpha$, $\beta$ and $C$ which are related by the following two conditions. The first one is the restriction imposed by the flatness condition and can be obtained from Eq. (\ref{eq6}) at $x=0$:
\be\label{eq9}
\frac{2}{3\beta-2\alpha+2}\Omega_{m0}+\frac{1}{2\beta-\alpha+1}\Omega_{r0}+C=1
\ee
The second condition is obtained by considering the equation of state for the present epoch values of the density and pressure (i.e. at x=0) of the dark energy $\tilde{p}_{\Lambda0}=\omega_0\Omega_{\Lambda0}$ (note that $\tilde{\rho}_{\Lambda0}=\Omega_{\Lambda0}$)
\be\label{eq10}
\frac{2\alpha-3\beta-2}{3\beta}\,C+\frac{2\beta-\alpha}{3(\alpha-2\beta-1)}\,
\Omega_{r0}=\omega_0\left[\frac{3\beta-2\alpha}{2\alpha-3\beta-2}\Omega_{m0}
+\frac{2\beta-\alpha}{\alpha-2\beta-1}\Omega_{r0}+C\right]
\ee
Solving Eqs. (\ref{eq9}) and (\ref{eq10}) with respect to $\beta$ it is obtained
\begin{equation}\label{eq11}
\alpha=\frac{1}{2}\left[2(1-\Omega_{m0}-\Omega_{r0})+\beta(\Omega_{r0}+3\omega_0(1-\Omega_{m0}-\Omega_{r0})+3)\right]
\end{equation}
and
\begin{equation}\label{eq12}
\begin{aligned}
C=&1-\frac{2\Omega_{m0}}{2(\Omega_{m0}+\Omega_{r0})-\beta\left[\Omega_{r0}+3\omega_0(1-\Omega_{m0}-\Omega_{r0})\right]}\\
&-\frac{2\Omega_{r0}}{2(\Omega_{m0}+\Omega_{r0})-\beta\left[\Omega_{r0}+3\omega_0(1-\Omega_{m0}-\Omega_{r0})-1\right]}
\end{aligned}
\end{equation}
Replacing this expressions for $\alpha$ and $C$ in (\ref{eq6}) we obtain a $\beta$-dependent Hubble parameter and cosmological evolution for a given matter density parameter $\Omega_{m0}$ and present dark energy EOs parameter $\omega_0$.
In previous works (\cite{granda,granda3,granda4}) we illustrated the behavior of the model by choosing some representative values for $\alpha$ and $\beta$ guided by the redshift transition in the deceleration parameter and found the evolution of equation of state $\omega(z)$. In general the obtained cosmological dynamics was very close to what is expected from the current astrophysical observations.
We turn now to use several data sets to constrain the parameter $\beta$ (and therefore $\alpha$) of the holographic dark energy model (\ref{eq2}). The constraints on the parameters of the holographic Ricci dark energy have been performed in \cite{zhang}.

\section{Constrains from SNIa observations}
We used the 307 super nova SNIa data from the union compilation set (table 11 from \cite{kowalski}). This 307 data are obtained after the cuts imposed on the total of 414 SNIa data using criteria of quality  \cite{kowalski}. The data set gives the distance modulus $\mu_{obs}(z_i)$, and the theoretical (for a given  model) distance modulus is defined by
\be\label{eq13}
\mu_{th}(z_i)=5Log_{10} D_L(z_i)+\mu_0
\ee
where $\mu_0=42.38-5Log_{10}h$, $h$ is the Hubble constant $H_0$ in units of 100 km/s/Mpc and $D_L(z)=H_0 d_L(z)/c$. The luminosity distance times $H_0$ is given by
\be\label{eq14}
d_L(z)=(1+z)\int_0^z\frac{cdz'}{\tilde{H}(z',\theta)}
\ee
where $\tilde{H}(z,\theta)$ from Eq. (\ref{eq6}) in terms of $z$ is given by
\be\label{eq14a}
\tilde{H}(z,\theta)=\left[\frac{2}{3\beta-2\alpha+2}\Omega_{m0}(1+z)^3
+\frac{1}{2\beta-\alpha+1}\Omega_{r0}(1+z)^4+C(1+z)^{2(\alpha-1)/\beta}\right]^{1/2}
\ee
with $\theta\equiv(\beta,\Omega_m,\omega_0)$ (after replacing $\alpha$ and $C$ from (\ref{eq11},\ref{eq12}) with $\Omega_{r0}=0$), but we will choose the values of $\Omega_m$ and $\omega_0$ and constraint the constant $\beta$. Next we minimize the statistical $\chi^2$ function (which determines likelihood function of the parameters) of the model parameters for the SNIa data.
\be\label{eq15}
\chi^{2}_{SN}(\theta)=\sum^{307}_{i=1}\frac{\left(\mu_{obs}(z_i)-\mu_{th}(z_i)\right)^2}{\sigma_i^2}
\ee
The $\chi^2$ function can be minimized with respect to the $\mu_0$ parameter, as it is independent of the data points and the data set \cite{nesseris}. Expanding the Eq. (\ref{eq15}) with respect to $\mu_0$ yields
\be\label{eq16}
\chi^{2}_{SN}(\theta)=A(\theta)-2\mu_0B(\theta)+\mu^2_0 C
\ee
which has a minimum for $\mu_0=B(\theta)/C$, giving 
\be\label{eq17}
\chi^{2}_{SN,min}(\theta)=\tilde{\chi}^{2}_{SN}(\theta)=A(\theta)-\frac{B(\theta)^2}{C}
\ee
with 
\be\label{eq18}
\begin{aligned}
A(\theta)=&\sum^{307}_{i=1}\frac{\left(\mu_{obs}(z_i)-\mu_{th}(z_i,\mu_0=0)\right)^2}{\sigma_i^2}\,\\
&B(\theta)=\sum^{307}_{i=1}\frac{\mu_{obs}(z_i)-\mu_{th}(z_i,\mu_0=0)}{\sigma_i^2}\,\\
&C=\sum^{307}_{i=1}\frac{1}{\sigma_i^2}
\end{aligned}
\ee
Now this $\tilde{\chi}^{2}_{SN}$ is independent of $\mu_0$ and can be minimized with respect to the parameters of the theoretical model. In our analysis we will take some samples of $\Omega_m$ and $\omega_0$ from the observational data and minimize with respect to the remaining parameter $\beta$. The best fit value for $\beta$ with $1\sigma$ uncertainty and the corresponding $\chi^{2}_{min}$, from the analysis of the Union sample of 307 SNIa \cite{kowalski} is resumed in table I.\\
\begin{table}[h]
	\centering
		\begin{tabular}{|c|c|c|c|c|c|}
		\hline

$\Omega_{m0}$ & $\omega_0$ & $\beta (1 \sigma)$          & $\alpha$   & h          & $\chi^{2}_{min}$ \\ \hline \hline
$0.28$     & $  -1.29$  & $0.292^{+0.040}_{-0.035}$  & $0.751$ & $0.708$ & $312.131       $ \\ \hline 
$0.28$     & $ -1$      & $0.584^{+0.146}_{-0.106}$    & $0.965$  & $0.698$ & $312.744        $\\ \hline    
$0.28$     & $-0.969$   & $0.655^{+0.185}_{-0.129}$  & $1.017$  & $0.696$ & $313.189        $\\ \hline

$0.29$     & $-1.29$    & $0.312^{+0.044}_{-0.038}$  & $0.749 $& $0.707$ & $312.035        $\\ \hline
$0.29$     &  $-1 $     & $0.636^{+0.167}_{-0.120}$  & $0.986 $& $0.697$ & $312.931        $\\ \hline
$0.29$     &  $-0.969$  & $0.717^{+0.214}_{-0.146}$  & $1.046 $ & $0.696$ & $313.414        $\\ \hline

$0.3$      &  $-1.29$   & $0.333^{+0.048}_{-0.041}$  & $0.748$ & $0.707$ & $311.951        $\\ \hline
$0.3$      &   $ -1$    & $0.693^{+0.193}_{-0.135} $ & $1.012   $ & $0.697$ & $313.138        $\\ \hline
$0.3$      &  $-0.969$  & $0.787^{+0.250}_{-0.167} $ & $1.080  $& $0.695$ & $313.659        $\\ \hline

		\end{tabular}
  	\caption{\it The best-fit values for $\beta$ with $1\sigma$ error, from the SN Ia analysis.}
\end{table}

The best fit value for $h$ can be obtained from the relation $\mu_0=B(\beta)/C$ at the best fit value for $\beta$. We used three representative values of $\Omega_{m0}$ combined with three representative values of $\omega_0$, obtaining a total of 9 best fit values for $\beta$. Note that $\beta$ significantly changes with the change in $\omega_0$ and the best fit $h$ is less sensitive, taking values in the region $h\sim 0.7$.
Fig. 1 shows the behavior of the Hubble parameter, using the SN Ia 307 data set, for three different best-fit values of $\beta$ taken from table I, and the corresponding likelihood behavior for the same three values.\\
\begin{figure}[hbtp]
\includegraphics[scale=0.85]{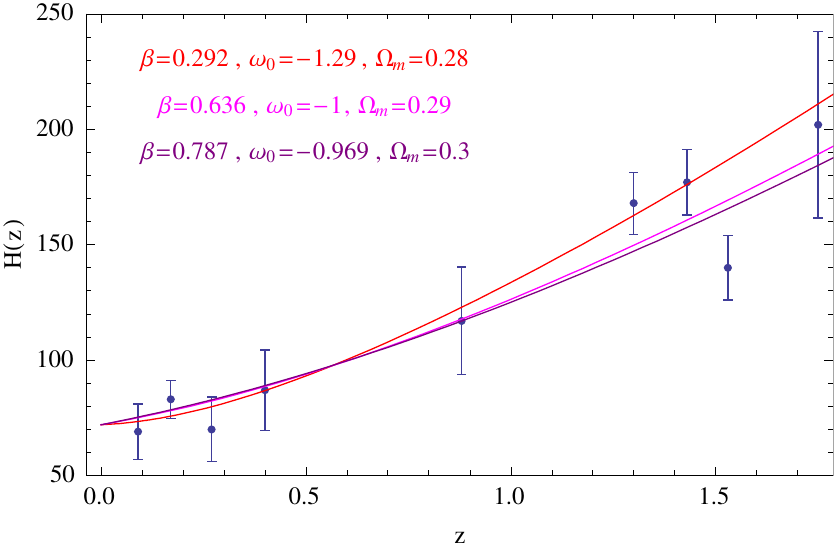}\hspace{0.5cm}\includegraphics[scale=0.85]{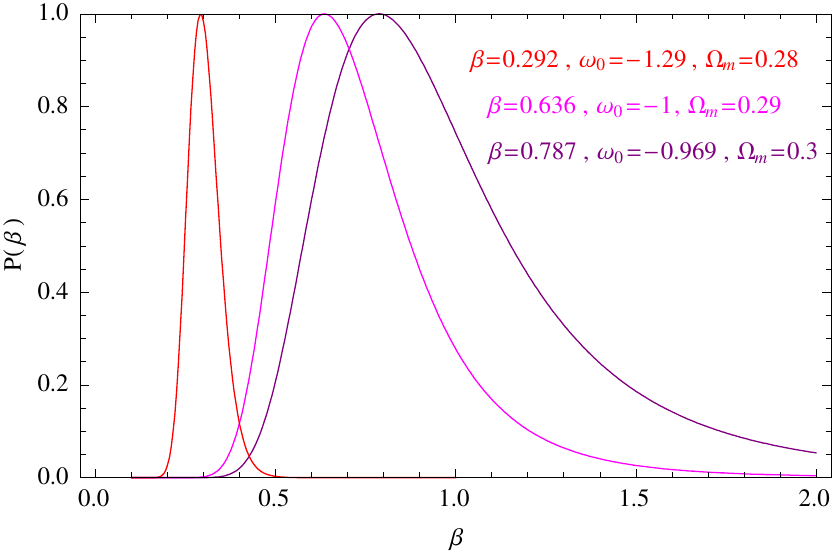}
\caption{\it The observational $H(z)$ data with error bars, and the holographic model $H(z)$ for three different best fit values of $\beta$ from the 307 SN Ia analysis, and (right) the corresponding likelihood behavior for the same set of parameters.}
\end{figure}

Note from table I, that in all cases for $\omega_0>-1$, the constant $\alpha>1$ giving a quintessence character to the model, as the power of $(1+z)$ in the last term in Eq. \ref{eq14a} becomes positive; and in the cases with $\omega_0<-1$, $\alpha$ changes to $\alpha<1$ and the power of $(1+z)$ becomes negative giving a quintom-like character to the model.
\newpage
\section{Constraining from the SNIa+CMB+BAO}
In this section we constrain $\beta$ in the holographic dark energy model (\ref{eq2}) by using the 
SNIa combined with the CMB and BAO observational data. The shift parameter $R$ \cite{bond} from the cosmic microwave background (CMB) anisotropy, and the distance parameter $A$ of the measurement of the baryon acoustic oscillation (BAO) peak in the distribution of SDSS luminous red galaxies \cite{tegmark}, are also used extensively in obtaining the cosmological constraints. The shift parameter $R$ is defined by 
\be\label{eq19}
R=\Omega^{1/2}_{m0}\int^{1090}_{0}\frac{dz}{\tilde{H}(z)}
\ee
where $z=1090$ is the redshift of the recombination \cite{komatsu}. The distance parameter $A$ is given by \cite{einsenstein}
\be\label{eq20}
A=\Omega^{1/2}_{m0}\tilde{H}(z_b)^{-1/3}\left[\frac{1}{z_b}\int^{z_b}_{0}\frac{dz}{\tilde{H}(z)}\right]^{2/3}
\ee
with $z_b=0.35$. And now, using the combined data of the 307 Union SN Ia, the shift parameter
$R$ of CMB and the distance parameter $A$ of BAO, we perform the joint analysis to constraint the constant $\beta$. The total $\chi^2$ is given by 
\be\label{eq21}
\chi^{2}=\tilde{\chi}^{2}_{SN}+\chi^{2}_{CMMB}+\chi^{2}_{BAO}
\ee
The best-fit model parameter can be determined by minimizing the total $\chi^{2}$. Here $\chi^{2}_{CMB}$  and $\chi^2_{BAO}$ are given by
\be\label{eq22}
\chi^{2}_{CMB}=\frac{(R-R_{obs})^2}{\sigma^2_R},\,\,\,\chi^{2}_{BAO}=\frac{(A-A_{obs})^2}{\sigma^2_A}
\ee
The SDSS BAO measurement (\cite{einsenstein}) gives the observed value of \linebreak
$A = 0.469(n_s/0.98)^{-0.35} \pm 0.017$ with the spectral index $n_s$ as measured by WMAP5 \cite{komatsu}, taken to be $n_s = 0.960$. The value of the shift parameter $R$ has also been updated by WMAP5 \cite{komatsu} to be $1.710\pm 0.019$. The table II shows the best fit value for $\beta$ with $1\sigma$ uncertainty and the corresponding $\chi^{2}_{min}$, from the joint analysis of the Union sample of 307 SNIa, CMB and BAO observations. We used a combination of three different values for $\Omega_{m0}$ and $\omega_0$ respectively.\\
\begin{table}[htbp]
	\centering
		\begin{tabular}{|c|c|c|c|c|c|}
		\hline

$\Omega_m$ & $\omega_0$ & $\beta(1\sigma)$     & $\alpha$   & h          & $\chi^{2}_{min}$ \\ \hline \hline
$0.28$     & $  -1.29$  & $0.437^{+0.016}_{-0.015}$  & $0.767$ & $0.718$ & $321.028       $ \\ \hline 
$0.28$     & $ -1$      & $0.625^{+0.023}_{-0.023}$  & $0.983$ & $0.699$ & $312.828      $\\ \hline    
$0.28$     & $-0.969$   & $0.654^{+0.025}_{-0.024}$  & $1.017$  & $0.696$ & $313.196        $\\ \hline

$0.29$     & $-1.29$    & $0.433^{+0.016}_{-0.015}$  & $0.765 $& $0.715$ & $319.404        $\\ \hline
$0.29$     &  $-1 $     & $0.618^{+0.024}_{-0.023}$  & $0.979 $& $0.697$ & $313.134        $\\ \hline
$0.29$     &  $-0.969$  & $0.646^{+0.025}_{-0.024}$  & $1.012 $ & $0.695$ & $313.663        $\\ \hline

$0.3$      &  $-1.29$   & $0.430^{+0.023}_{-0.016}$  & $0.763$   & $0.713$ & $318.597        $\\ \hline
$0.3$      &   $ -1$    & $0.611^{+0.024}_{-0.023}$  & $0.975$ & $0.695 $ & $314.055        $\\ \hline
$0.3$      &  $-0.969$  & $0.638^{+0.023}_{-0.024}$  & $1.008  $& $0.693$ & $314.727        $\\ \hline

		\end{tabular}
  	\caption{\it The best-fit values for $\beta$ with $1\sigma$ error, from the joint SNIa+CMB+BAO analysis.}
\end{table}

Note that $R$ and $A$ are independent of $\mu_0$, but the best fit value for $h$ is obtained from $\mu_0=B(\beta)/C$ at the new best-fit values for $\beta$. Fig. 2 shows the behavior of $H(z)$ for the joint analysis of the SNIa+CMB+BAO data, and the likelihood  behavior for the SNIa+CMB+BAO analysis, for three representative best-fit values of $\beta$ taken from table II.
\begin{figure}[hbtp]
\includegraphics[scale=0.85]{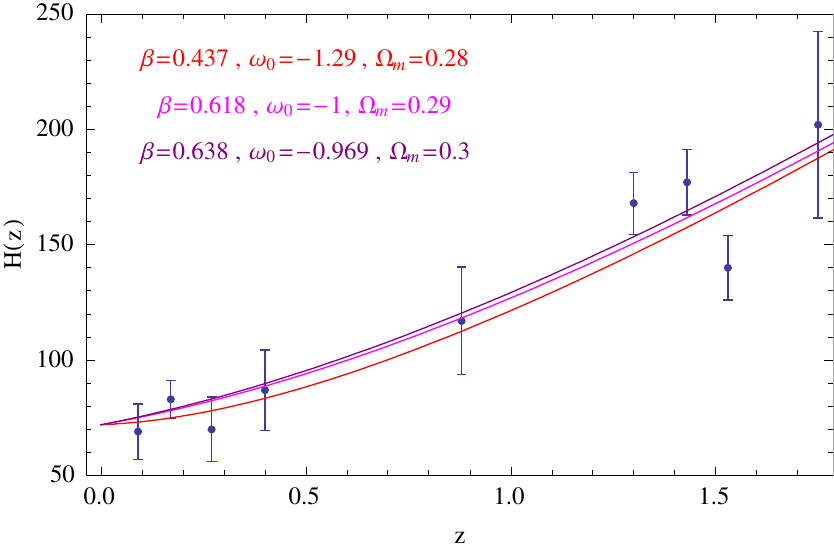}\hspace{0.5cm}\includegraphics[scale=0.85]{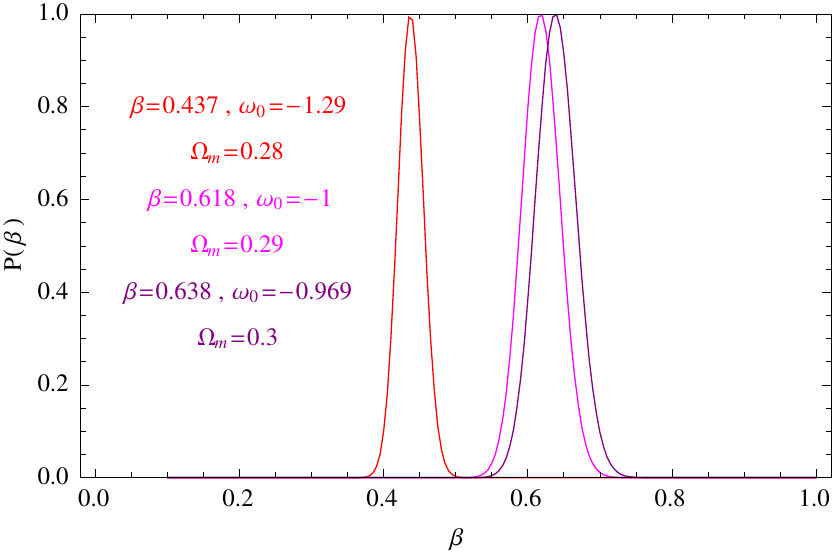}
\caption{\it The observational $H(z)$ data with error bars, and the holographic model $H(z)$ for three different best fit values of $\beta$ from the joint 307 SNIa+CMB+BAO analysis, and (right) the corresponding likelihood behavior for the same set of parameters.}
\end{figure}

\noindent Looking at tables I and II, and analyzing all the obtained values for $\chi^2_{min}$, the best fit for the SN Ia analysis happens at $\beta=0.333^{+0.048}_{-0.041}$, corresponding to the lowest $\chi^2_{min}=311.951$ with $\Omega_{m0}=0.3$ and $\omega_0=-1.29$, and for the joint SN Ia+CMB+BAO analysis, at $\beta=0.625^{+0.023}_{-0.023}$, lowest $\chi^2_{min}=312.828$ with $\Omega_{m0}=0.28$ and $\omega_0=-1$.
Fig. 3 shows the behavior of the Hubble parameter with the redshift for the lowest $\chi^2_{min}$ from the SN Ia analysis and the SNIa+CMB+BAO joint analysis, and the corresponding likelihood behavior.
\begin{figure}[hbtp]
\includegraphics[scale=0.85]{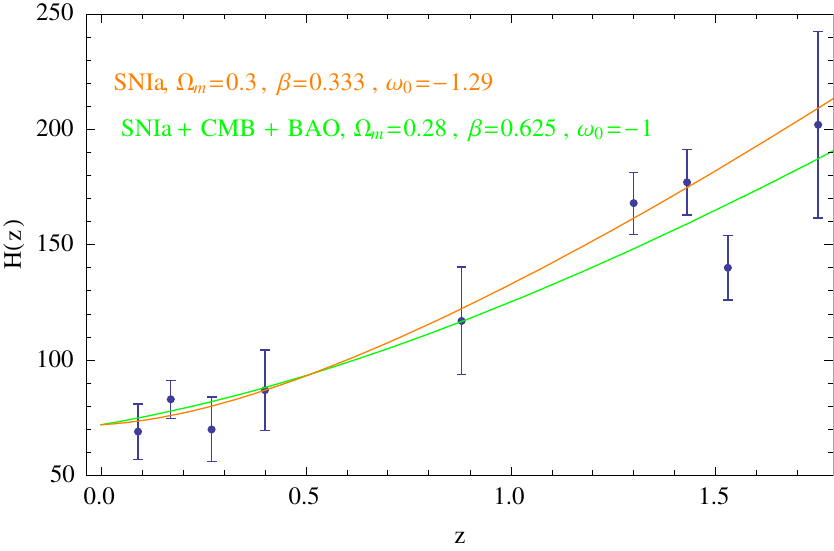}\hspace{0.5cm}\includegraphics[scale=0.85]{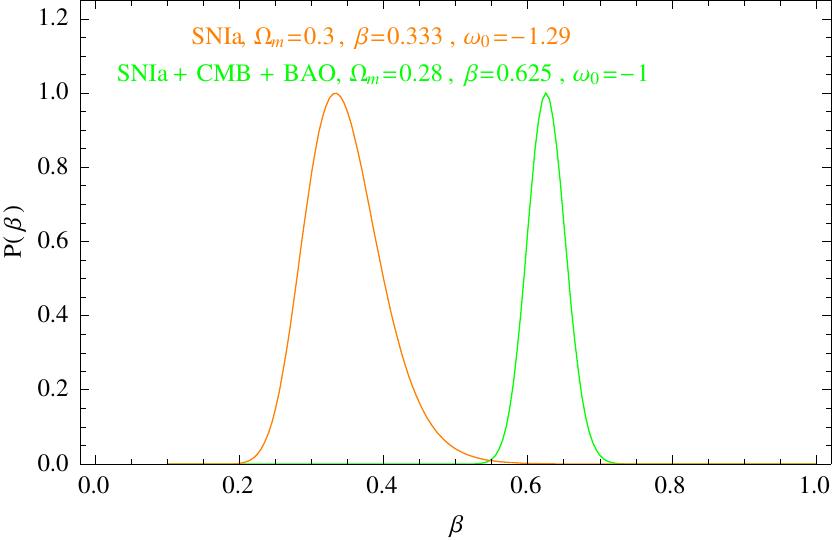}
\caption{\it The observational $H(z)$ data with error bars, and the holographic model $H(z)$ for the best-fit $\beta$ corresponding to the lower $\chi^2_{min}$ from the SNIa (table I) and the joint SNIa+CMB+BAO (table II) analysis, and (right) the corresponding likelihood behavior.}
\end{figure}

Note that we have constrained $\beta$ alone, fixing each time the current values of $\Omega_{m0}$ and $\omega_0$, and hence we can not make definite conclusion about the quintessence or quintom nature of the model (\ref{eq2}). Nevertheless looking at Fig. 3 for experimental (error bars) versus holographic model $H(z)$, it can be seen that the quintom behavior ($\alpha<1$) is favored by both, the SNIa and joint SNIa+CMB+BAO analysis.\\
To better illustrate the cosmological dynamics, in Fig. 4 we plot the effective equation of state parameter ($\omega_{eff}=p_{\Lambda}/(\rho_{\Lambda}+\rho_m$) and the deceleration parameter $q(z)$ for the two best-fit $\beta$ used in Fig. 3.
\begin{figure}[hbtp]
\includegraphics[scale=0.85]{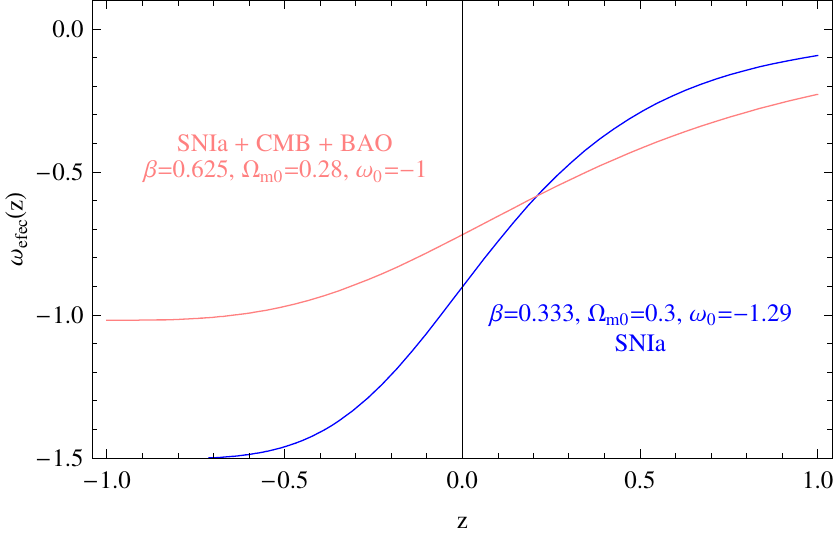}\hspace{0.5cm}\includegraphics[scale=0.85]{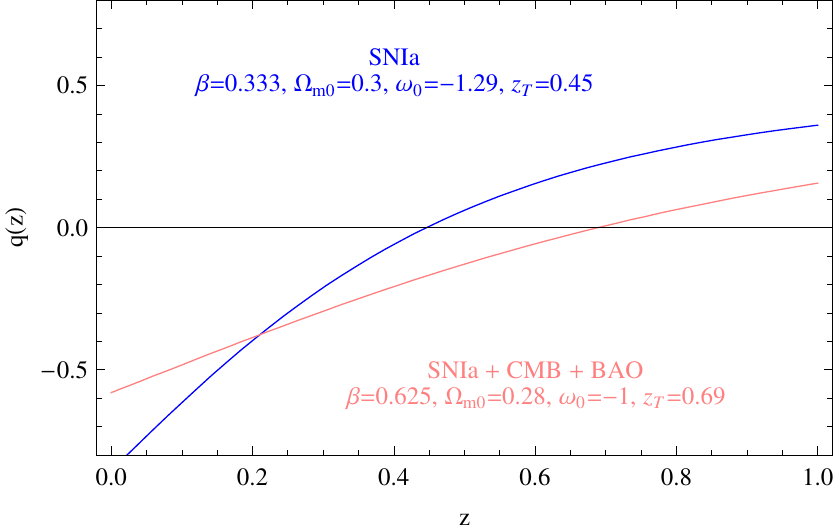}
\caption{\it The the evolution of the effective equation-of-state $\omega_{eff}(z)$, for the best-fit $\beta$ from the SNIa (table I) and the joint SNIa+CMB+BAO (table II) analysis, and (right) the deceleration parameter showing the redshift transition $z_T$ for both cases. The redshift transitions $z_T$ are in the range accepted by observations}
\end{figure}

\section{Constraining the cosmological and model parameters using Montecarlo}
In this section we perform constraints on the three parameters $\beta$, $\Omega_{m0}$ and $\omega_0$ of our holographic dark energy model, combining the observations from 307 SNIa, CMB and BAO, by using the montecarlo technique to restrict the $\chi^2_{min}$ and choosing an appropriate intervals for $\beta$, $\Omega_{m0}$ and $\omega_0$. The analysis was done over the total of $10^6$ data, with the restrictions on $\chi^2_{min}$ and $\beta$ taken according to the previous fits consigned in tables I and II: $\chi^2_{min}<313.000$, $0.25<\beta<2.3$, $0.25<\Omega_{m0}<0.35$ and $-1.3<\omega_0<-0.8$. After $10^6$ iterations, the best fit-value for the triplet of model parameters was $\beta=0.593^{+0.021}_{-0.023}(1\sigma)$, $\Omega_{m0}=0.283^{+0.014}_{-0.013}(1\sigma)$ and $\omega_0=-1.036^{+0.027}_{-0.024}(1\sigma)$, with $\chi^2_{min}=312.734$. The value of $\omega_0$ is slightly lower than $-1$ showing that the model tends to behave as quintom. Fig. 5 shows the Hubble parameter versus $z$ for the best-fit values of $(\beta,\Omega_{m0},\omega_0)$, and the evolution of the effective and holographic EoS parameters.
\begin{figure}[hbtp]
\includegraphics[scale=0.85]{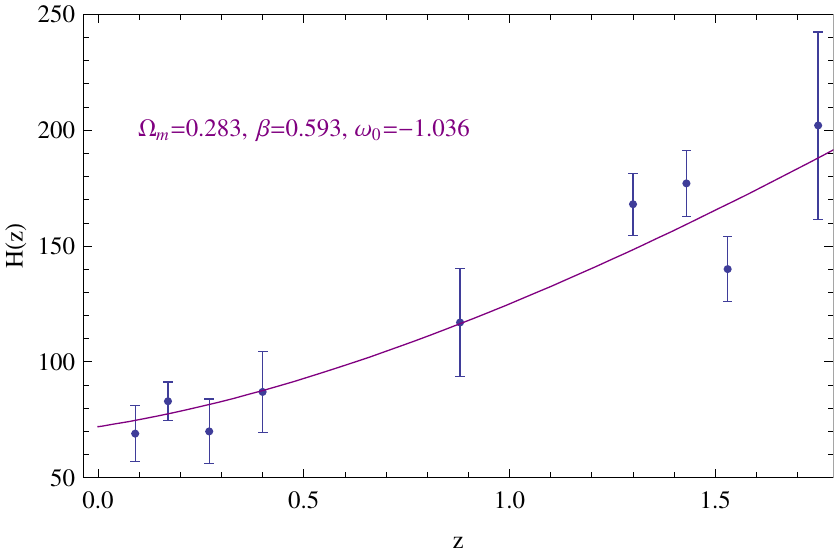}\hspace{0.5cm}\includegraphics[scale=0.85]{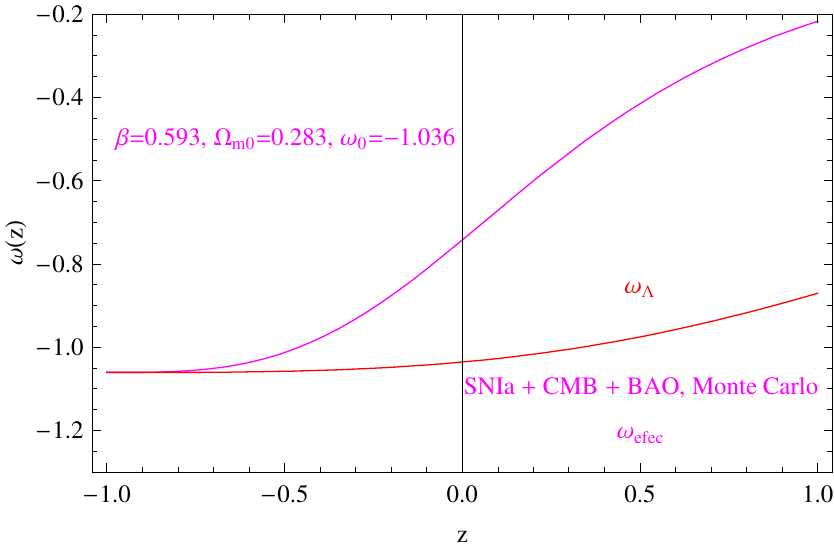}
\caption{\it The observational $H(z)$ data with error bars, and the holographic $H(z)$ for the best-fit model parameters $(\beta,\Omega_{m0},\omega_0)$. The right graphic shows the effective EoS and the holographic EoS $\omega_{\Lambda}(z)$. The transition deceleration acceleration ($\omega_{eff}=-1/3$) occurs at $z_T\approx0.67$. The quintom character of the model with the $\omega_{\Lambda}=-1$ crossing is seen from the $\omega_{\Lambda}$ behavior.}
\end{figure}

\section{The Statefinder and Om diagnostics}
To better understand the properties of the DE model (\ref{eq2}) is useful to compare with the model independent diagnostics which are able to differentiate between a wide variety of dynamical DE models, including the $\Lambda$CDM model \cite{sahni1}. We will use the diagnostic introduced in \cite{sahni1} known as statefinder, which introduces a pair of parameters ($r$, $s$) defined as
\be\label{eq23}
r=\frac{\dddot{a}}{aH^3},\ \ \ \ \ s=\frac{r-1}{3(q-1/2)}
\end{equation}
which are proven to be useful to characterize a given DE model, as this pair depend only on high derivatives of the scale factor $\ddot{a}$ and $\dddot{a}$, acquiring some geometrical sense as depend only on the spacetime metric. 
In terms of the total density and pressure ($r$, $s$) can be written as  (\cite{sahni}).
\be\label{eq24}
r=1+\frac{9(\rho+p)}{2\rho}\frac{\dot{p}}{\dot{\rho}},\ \ \ \ \ s=\frac{(\rho+p)}{p}\frac{\dot{p}}{\dot{\rho}}
\end{equation}
Is clearly seen from Eq. (\ref{eq24}) that in the flat FRW background, the $\Lambda$CDM model corresponds to a fixed point $(s=0, r=1)$ in the $r-s$ plane. The trajectories in the $s-r$ plane corresponding to different DE models may exhibit qualitatively different behaviors, and this is a good way to establish the departure of a given DE model from the $\Lambda$CDM. In this work we apply the statefinder diagnostic to the holographic DE model (\ref{eq2}) (\cite{granda}), using the calculated above best-fit parameters with the SNIa and joint SNIa+CMB+BAO data analysis.\\
\noindent The statefinder parameters for the model (\ref{eq2}) are given by 
\begin{equation}\label{eq25}
r=1+\frac{C(\alpha-1)(2-2\alpha+3\beta)^2e^{3x}}{\beta^2\left[C(2\alpha-3\beta-2)e^{3x}-2\Omega_{m0}e^{2x(\alpha-1)/\beta}\right]}
\end{equation}
and
\begin{equation}\label{eq26}
s=\frac{2}{3}\left(\frac{\alpha-1}{\beta}\right)
\end{equation}
Note that $\alpha$ and $C$ have to be replaced by solutions (\ref{eq11}) and (\ref{eq12}). From this Equations one can see that $s=0$ and $r=1$ for $\alpha=1$, which corresponds to the $\Lambda$CDM model. To illustrate our case, in Fig. 6 we plot the statefinder diagram in the $s-r$ plane for the best-fit $(\beta,\Omega_{m0},\omega_0)$ taken from the joint SNIa+CMB+BAO analysis using montacarlo and for the best-fit $\beta=0.654, \alpha=1.017$ ($\Omega_{m0}=0.28, \omega_0=-0.969$) taken from table II, for comparison.
\begin{center}
\includegraphics[scale=0.7]{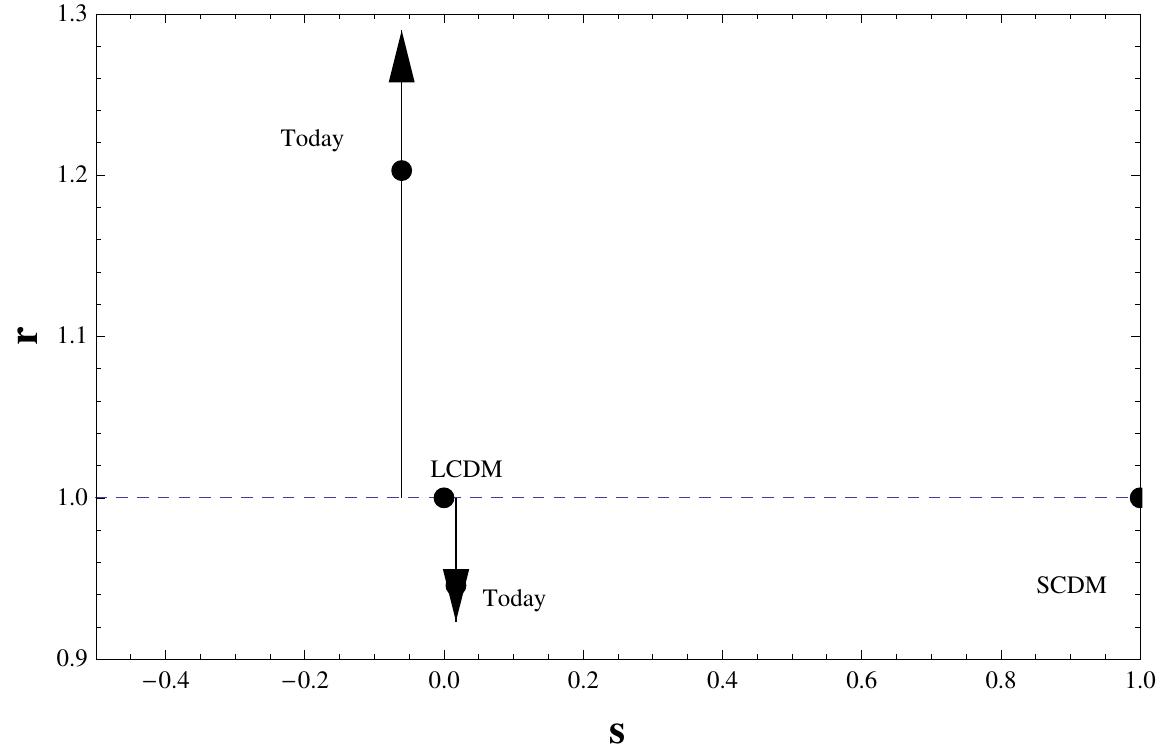}
\end{center}
{\it Figure 6: The statefinder evolution for the holographic model in the $s-r$ plane with ($\beta=0.593$, $\alpha=0.946$, $C=0.7$, $\Omega_{m0}=0.283$)(arrow up) and ($\beta=0.654$, $\alpha=1.017$, $C=0.71$, $\Omega_{m0}=0.28$) (arrow down)}.

Since $s$ is constant, the trajectory in the $s-r$ plane is a vertical line with $r$ monotonically increasing from $r=1$ to $r\approx 1.29$ (at $x->\infty$ in (\ref{eq25})) for the used best-fit values. The current $(s,r)$ values are $(s=-0.06,  r(x=0)=1.203)$. The SCDM (standard cold dark matter) model is shown in the point $(s=1, r=1)$. The evolution trajectory in the $s-r$ plane is similar to the one obtained for the Ricci DE model \cite{jun}. Note that for $\alpha>1$ the $r$ trajectory points in the opposite direction.\\
Another useful statefinder diagram is given by the trajectory in the $q-r$ plane. The deceleration parameter $q=\frac{1}{2}\left(1+\frac{3p}{\rho}\right)$ applied to the model (\ref{eq2}) is given by
\begin{equation}\label{eq27}
q=\frac{1}{2}\left[1+\frac{C(2-2\alpha+3\beta)^2e^{3x}}{\beta\left[C(2\alpha-3\beta-2)e^{3x}-2\Omega_{m0}e^{2x(\alpha-1)/\beta}\right]}\right]
\end{equation}
and $r$ in terms of $q$ is
\begin{equation}\label{eq28}
r=1+\left(\frac{\alpha-1}{\beta}\right)(2q-1)
\end{equation}
Fig. 7 shows the evolution trajectory of the model in the $q-r$ plane for the same best-fit values of fig. 6. 
\begin{center}
\includegraphics[scale=0.7]{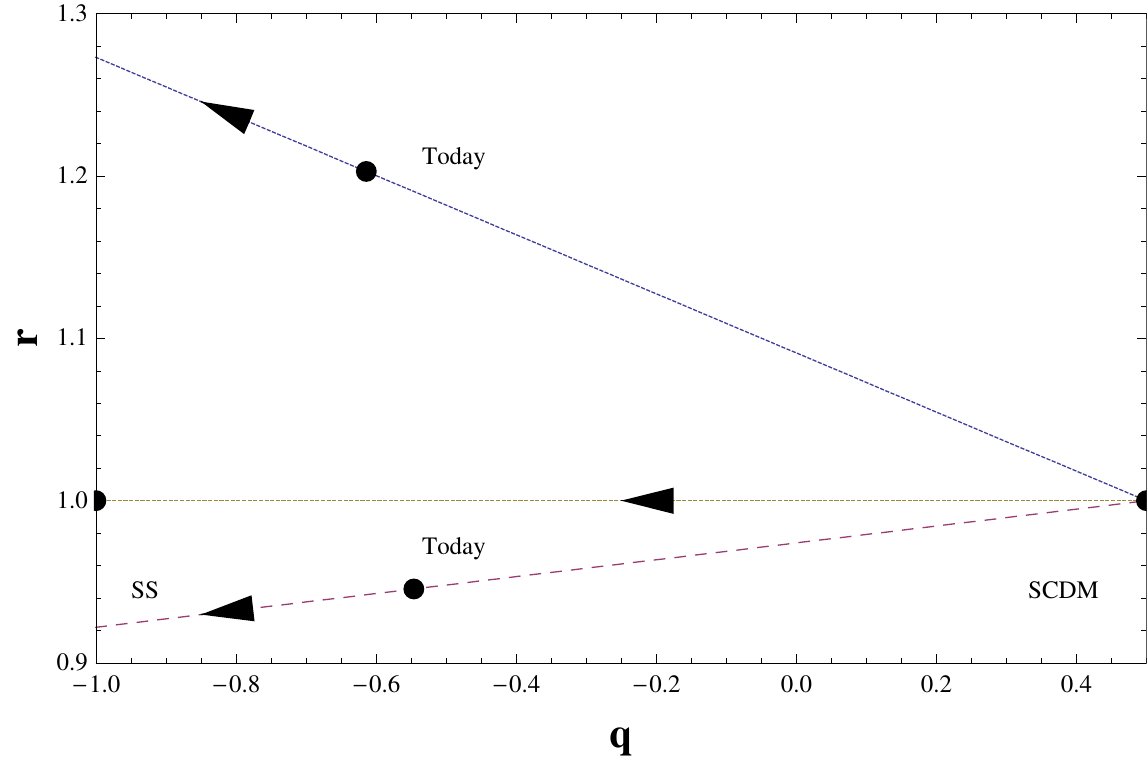}
\end{center}
{\it Figure 7: The statefinder evolution for the holographic model in the $q-r$ plane, for ($\beta=0.593$, $\alpha=0.946$, $C=0.7$, $\Omega_{m0}=0.283$}) and ($\beta=0.654$, $\alpha=1.017$, $C=0.71$, $\Omega_{m0}=0.28$).

we see that the $\Lambda$CDM and the holographic model start diverging from the same point in the past (the standard cold dark matter SCDM). The current values of the statefinder parameters are in $q=-0.614$ and $r=1.203$. This trajectory has constant negative slope, corresponding to a quintom behavior of the model with $\alpha<1$. The trajectory for $\alpha=1.07$ taken from table II (for illustration) have positive slope as can also be seen from Eq. (\ref{eq28}) for $\alpha>1$, characterizing the quintessence behavior of the proposed holographic DE model. In this plane the $\Lambda$CDM scenario is the horizontal line in Fig. 7 (as follows from Eq. (\ref{eq28})) , and evolves from the SCDM in the past ($q=0.5, r=1$), ending at the steady state cosmology (SS) in the future ($q=-1, r=1$).\\
An alternative way of distinguishing $\Lambda$CDM from other DE models without directly involving the EoS, is the $Om$ diagnostic introduced in \cite{sahni2}. This $Om$ diagnostic is constructed from the Hubble parameter, depending only on the first derivative of the expansion factor $a(t)$, and hence depends only upon the expansion history of our Universe. It is defined by 
\begin{equation}\label{eq29}
Om(x)=\frac{\tilde{H}^2(x)-1}{e^{-3x}-1}
\end{equation}
which for the present holographic model turns out to be
\begin{equation}\label{eq30}
Om(x)=\frac{\frac{2\Omega_{m0}}{2-2\alpha+3\beta}e^{-3x}+Ce^{-2x(\alpha-1)/\beta}-1}{e^{-3x}-1}
\end{equation}
for DE models with constant EoS the $Om(x)$ takes a simple form (see \cite{sahni2}) and for the $\Lambda$CDM model takes the simplest form, being equal to the density parameter \cite{sahni2}. In Fig. 8 we show the $Om$ diagnostic for the holographic model with the best-fit $(\beta,\Omega_{m0},\omega_0)$ taken from the joint SNIa+CMB+BAO analysis using montacarlo, and for the best-fit $\beta=0.654, \alpha=1.017$ ($\Omega_{m0}=0.28, \omega_0=-0.969$) taken from table II for comparison.
\begin{center}
\includegraphics[scale=0.7]{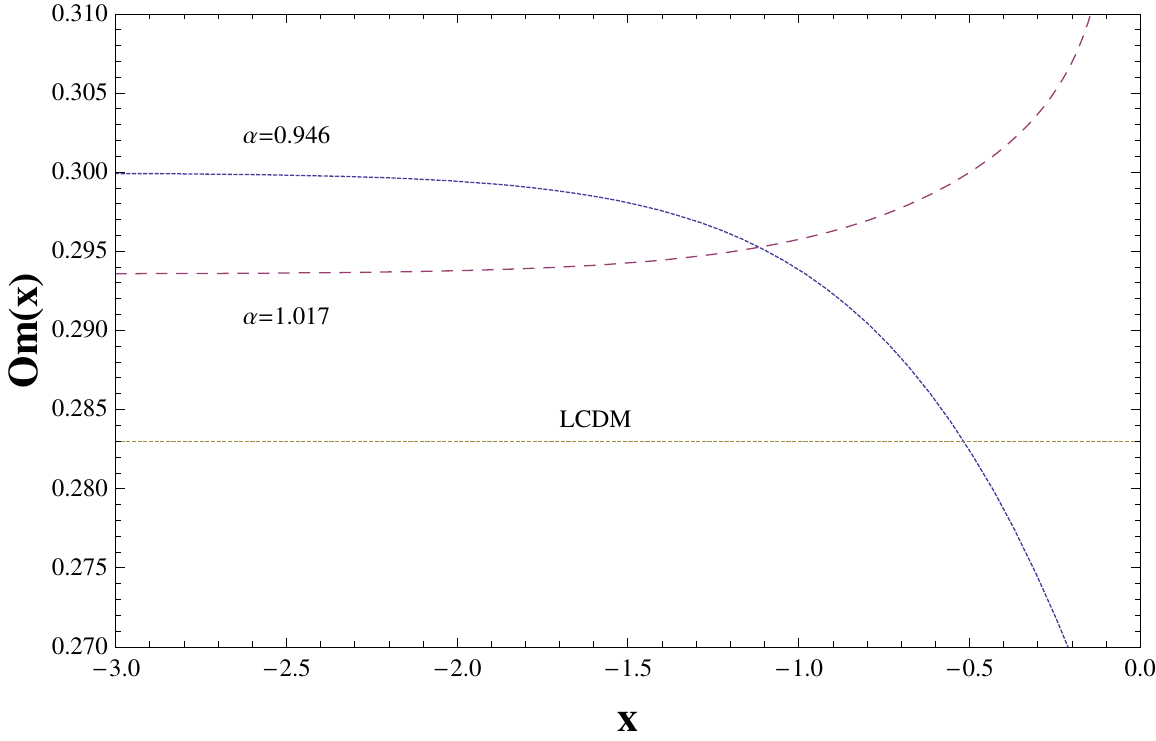}
\end{center}
{\it Figure 8: The $Om(x)$ for the holographic model with the montecarlo best-fit $\beta=0.593$, $\alpha=0.946$, $C=0.7$ and $\Omega_{m0}=0.283$, and for the best-fit $\beta=0.654, \alpha=1.017$ from table II.}

In this graphic the evolutionary difference between the $\Lambda$CDM and the studied holographic DE model is more clear, with the $Om$ curve turning down for $\alpha<1$ and up for $\alpha>1$.

\section{discussion}
In this letter we have obtained the constraints on the parameters of the holographic dark energy model described by the density Eq. (\ref{eq2}) \cite{granda}, using the latest observational data including the 307 Union sample of SNIa, the CMB shift parameter given by WMAP5, and BAO measurement from SDSS. We first used the 307 Union sample of SNIa to constraint the parameter $\beta$, assuming priors about fundamental cosmological quantities such as the current EoS for dark energy and the matter density parameter, and have generated the table I with different priors proposed. Looking at the minimum of the different $\chi^2_{min}$ listed in table I, a representative best-fit can be given by $\beta=0.333^{+0.048}_{-0.041}$ ($1\sigma$) with priors $\Omega_{m0}=0.3$ and $\omega_0=-1.29$. The same considerations for the joint  SNIa+CMB+BAO analysis resumed in table II, give the best-fit $\beta=0.625^{+0.023}_{-0.023}$ with priors $\Omega_{m0}=0.28$ and $\omega_0=-1$, with $1\sigma$ uncertainty. In fig. 3 we plot the evolution of $H(z)$ for the two best-fit $\beta$ for each analysis. Of course, the assumed priors may subvert the efficacy of the method and would be better to minimize the $\chi^2$ with respect to all the relevant cosmological parameters. 
In this sense we considered useful to use the montecarlo method to generate the $\chi^2_{min}$ without assuming priors about the relevant parameters, but restricting their range to a reasonable intervals (between the limits on $\Omega_{m0}$ and $\omega_0$ set by experiments). After $10^6$ iterations we obtained the following best fits with $1\sigma$ uncertainty: $\beta=0.593^{+0.021}_{-0.023}$, $\Omega_{m0}=0.283^{+0.014}_{-0.013}$ and $\omega_0=-1.036^{+0.027}_{-0.024}$, with $\chi^2_{min}=312.734$. The value of $\omega_0$ states that the possibility of quintom behavior can not be discounted. The behavior of H(z) contrasted with the observational error bars is shown in Fig. 5. Note also that the values of best-fit $h$ were found in a very narrow intervals (see tables I and II): for the 307 SNIa data $0.695<h<0.708$ and for the joint SNIa+CMB+BAO $0.697<h<0.718$, which are in the limits set by different experiments \cite{spergel}, \cite{tegmark2}, \cite{freedman}. From Figs. 3 and 5 for $H(z)$ and using the standard deviation from the observational error bars, the best-fit curve using the standard deviation criteria was obtained for the montacarlo SNIa+CMB+BAO analysis (Fig. 5).\\
\noindent we also used the statefinder and $Om$ diagnosis to differentiate this model with other models of dark energy. With the statefinder diagnosis the evolutionary trajectory in both, the $s-r$ and $q-r$ planes points in opposite directions depending on the quintessence ($\alpha>1$) or quintom ($\alpha<1$) nature of the model. In the $q-r$ plane the trajectory diverges from the SCDM point, but never reaches the $\Lambda$CDM line (except for $\alpha=1$), making the difference with other DE models \cite{sahni1}, \cite{xin}. With the $Om$ diagnosis there is a marked difference between the holographic model and the $\Lambda$CDM at the current time, and in the case of $\alpha<1$ the holographic trajectory intercepts in the past (at $z\sim 0.65$) the $\Lambda$CDM line. Finally we hope that future high precision
experiments allow accurately determine the parameters of the model to define its quintessence or quintom nature.

\section*{Acknowledgments}
This work was supported by the Universidad del Valle.

\end{document}